# Using association rule mining and ontologies to generate metadata recommendations from multiple biomedical databases

Marcos Martínez-Romero* <marcosmr@stanford.edu>, ORCID: 0000-0002-9814-3258

Martin J. O'Connor <martin.oconnor@stanford.edu>, ORCID: 0000-0002-2256-2421

Attila L. Egyedi <egyedia@stanford.edu>, ORCID: 0000-0003-0730-5053

Debra Willrett <willrett@stanford.edu>, ORCID: 0000-0002-3767-2957

Josef Hardi <johardi@stanford.edu>, ORCID: 0000-0002-2533-6681

John Graybeal <jgraybeal@stanford.edu>, ORCID: 0000-0001-6875-5360

Mark A. Musen <musen@stanford.edu>, ORCID: 0000-0003-3325-793X

Stanford Center for Biomedical Informatics Research, 1265 Welch Road, Stanford University School of Medicine, Stanford, CA 94305-5479, USA.

* Corresponding author

# Abstract

Metadata—the machine-readable descriptions of the data—are increasingly seen as crucial for describing the vast array of biomedical datasets that are currently being deposited in public repositories. While most public repositories have firm requirements that metadata must accompany submitted datasets, the quality of those metadata is generally very poor. A key problem is that the typical metadata acquisition process is onerous and time consuming, with little interactive guidance or assistance provided to users. Secondary problems include the lack of validation and sparse use of standardized terms or ontologies when authoring metadata. There is a pressing need for improvements to the metadata acquisition process that will help users to enter metadata quickly and accurately. In this paper we outline a recommendation system for metadata that aims to address this challenge. Our approach uses association rule mining to uncover hidden associations among metadata values and to represent them in the form of association rules. These rules are then used to present users with real-time recommendations when authoring metadata. The novelties of our method are that it is able to combine analyses of metadata from multiple repositories when generating recommendations and can enhance those recommendations by aligning them with ontology terms. We implemented our approach as a service integrated into the CEDAR Workbench metadata authoring platform, and evaluated it using metadata from two public biomedical repositories: US-based National Center for Biotechnology Information (NCBI) BioSample and European Bioinformatics Institute (EBI) BioSamples. The results show that our approach is able to use analyses of previous entered metadata coupled with ontology-based mappings to present users with accurate recommendations when authoring metadata.

# 1  Introduction

In the past decade there has been an explosion in the number of biomedical datasets submitted to public repositories, primarily driven by the requirement of journals and funding agencies to make experimental data openly available (1). Publicly funded organizations, such as the US-based National Center for Biotechnology Information (NCBI) and the European Bioinformatics Institute (EBI), have met this need by developing an array of repositories that allow the dissemination of datasets in the life sciences. These repositories typically impose detailed restrictions on the metadata that must accompany submitted datasets, generally driven by metadata specifications called minimum information models (2). The availability of these descriptive metadata is critical for facilitating online search and informed, secondary analysis of experimental results.

Despite the strong focus on requiring rich metadata for dataset submissions, the quality of the submitted metadata tends to be extremely poor (3,4). A significant problem is that creating well-specified metadata takes time and effort, and scientists view metadata authoring as a burden that does not benefit them (5). A typical submission requires spreadsheet-based entry of metadata—with metadata frequently spread over multiple spreadsheets—followed by manual assembly of multiple spreadsheets and raw data files into an overall submission package. Further problems occur because submission requirements are typically written at a high level of abstraction. For example, while a standard may require indicating the organism associated to a biological sample, it typically will not specify how the value of the organism must be supplied. Little use is made of the large number of ontologies available in biomedicine. Submission processes reflect this lack of precision ensuring that unconstrained, string-based values become the norm. Weak validation further exacerbates the problem, leading to metadata submissions that are sparsely populated and that frequently contain erroneous values (4).

In this paper, we describe the development of a method and associated tools that aim to address this quality deficit in metadata submissions. A central focus of this work is to accelerate the metadata acquisition process by providing recommendations to the user during metadata entry and—when possible—to help increase metadata adherence to the FAIR principles (6) by presenting recommendations that correspond to the most suitable controlled terms. Our method uses a well-established data mining technique known as association rule mining to generate real-time suggestions based on analyses of previously entered metadata.

## 2  Related Work

Web browsers have provided auto-fill and auto-complete functionality since the early days of the Web. Typical auto-fill functionality includes the automatic population of address and payment fields when completing Web-based forms. Auto-complete suggestions are primarily provided during page URL entry, usually driven by a simple frequency-based analysis of previously visited pages. Significantly more advanced auto-fill functionality is provided by Web search vendors, where suggestions are based both on in-depth analyses of Web content and on previous user behavior (7).

More specialized recommendations systems have also been developed to assist users when completing general-purpose fields in custom forms. Instead of concentrating on commonly predicted fields, such as ZIP code, the goal is to provide auto-fill and auto-complete functionality for as many fields as possible. These systems generally generate recommendations based on analyses of previously completed forms. An example is Usher (8), which was developed to speed-up form completion by analyzing previous submissions of the same form to predict likely values for fields during completion of a new form. A similar system called iForm (9) provides suggestions by learning field values from previously submitted forms. More advanced approaches provide predictive capabilities by combining analyses from multiple structurally different forms used in the same application domain. A system called Carbon (10), for example, uses a semantic mapping process to align fields in different forms so that values from existing distinct forms can be used to present auto-complete suggestions for fields in a new form.

Further predictive enhancements are possible when using *context-based* methods, which generate successively more refined field–value predictions as more fields are filled in. Instead of predicting field values on a form in isolation, these approaches refine their predictions by considering the values of other form fields that have already been populated. One of the earliest context-based systems was described by Ali and Meek (11). This system uses a predictive model to generate recommendations for a field by combining field-level analyses from previously completed forms with the context provided by fields that have already been populated.

Other systems have focused on presenting suggestions for field values that represent terms from controlled terminologies. Instead of allowing users to fill in free text for field values, these systems present auto-fill and auto-complete suggestions that correspond to terms in ontologies and controlled terminologies. Systems that provide this capability include RightField (12), ISA-tools (13), and Annotare (14). These systems have been used to increase metadata quality in the biomedical domain. None of these systems provide recommendations based on analyzing previously entered values, however.

In the context of the existing literature, our method belongs to a category of supervised learning models known as associative classifiers. This technique was first published in 1998 (15), and it has been broadly investigated and exploited by the data mining and machine learning communities in a number of successful real-word applications (16–19). Associative classifiers use association rule mining to extract interesting rules from the training data, and the extracted rules are used to build a classifier. In this work, the resulting classifier is used to predict field values. Associative classifiers usually scale well and have the advantage that the generated rules are meaningful, easy to interpret, and easy to debug and validate by domain experts. Several works have shown that associative classifiers often produce more accurate results than traditional classification techniques (20–23).

The work outlined in this paper is the first recommendation approach that combines the ability to offer predictions based on context-based methods with the standardization capabilities of ontology-based suggestions. Crucially, our approach extends our initial research (24) by enabling recommendations based on values entered for multiple metadata-acquisition forms, which may have different structure and field names. In this paper we outline our method and present an implementation.

## 3   Methods

We have designed an approach that uses association rule mining to discover hidden patterns in the values entered for fields in electronic forms. These patterns are then used to recommend the most appropriate choices when entering new field values. The approach works for both plain text values and ontology-based values. We first outline our method and then explain how we implemented it in the CEDAR Workbench.

## 3.1 Description of the approach

Our approach is centered on the notion of *templates*. A template defines a set of data attributes, which we call *template fields* or *fields*, that users fill in with values. For example, an *Experiment* template may have a *sex* field to enter the physical sex of the sampled organism (e.g., *female*), a *tissue* field to capture the type of tissue tested in the experiment (e.g., *skin*), and a *disease* field to enter the disease of interest (e.g., *psoriasis*).

Every time a template is filled in with values, a new *template instance* (or *instance*) is created. Given $F = \{f_1, f_2, \ldots, f_n\}$ a set of non-empty fields and $V = \{v_1, v_2, \ldots, v_n\}$ the set of values assigned to the fields in $F$, we define an instance $I_i$ as:

$$I_i = \{(f_1, v_1), \ldots, (f_n, v_n)\}$$

That is, an instance is a set of field–value pairs $(f_i, v_i)$, such that the value entered for the field $f_i$ is $v_i$. We will also represent a field–value pair as $f_i = v_i$. An example of instance of the *Experiment* template is as follows:

$$\{(sex, male), (tissue, liver), (disease, liver\ cancer)\}$$

This instance can also be represented as:

$$(sex = male) \wedge (tissue = liver) \wedge (disease = liver\ cancer)$$

We define an *instance repository I*, which stores all the template instances that have been created, as follows:

$$I = \{I_1, I_2, \ldots, I_n\}$$

where $I_i$ are template instances. Each instance is derived from one and only one template. Typically, an instance repository contains instances for a variety of different templates. Given $T$ the set of all templates, we define a function $template: I \rightarrow T$ that returns the template that the instance instantiates. Table 1 shows the content of an example repository that contains six instances for the *Experiment* template.

Table 1. Content of an example repository with six instances of an *Experiment* template. For each instance, the table shows its fields and the values assigned to them. Fields with empty values are omitted.

| Instance | Field–Value pairs |
|---|---|
| $I_1$ | $(sex = male) \wedge (tissue = brain) \wedge (disease = meningitis)$ |
| $I_2$ | $(sex = female) \wedge (tissue = brain) \wedge (disease = meningitis)$ |
| $I_3$ | $(tissue = liver) \wedge (disease = cirrhosis)$ |
| $I_4$ | $(sex = male) \wedge (tissue = liver) \wedge (disease = liver\ cancer)$ |
| $I_5$ | $(tissue = liver) \wedge (disease = liver\ cancer)$ |
| $I_6$ | $(sex = male) \wedge (tissue = brain) \wedge (disease = meningitis)$ |

Users add new instances to the repository by populating templates, that is, by entering values for template fields. A key focus of our method is considering the values of previously populated fields in the current instance when suggesting values for an active field. We refer to these existing values as the *context* and refer to the active field as the *target field*.

The concept of *context* is crucial, as we believe that the contextual information given by previously entered values can be used to suggest the most appropriate value for the target field. Formally, we define the context $C$ of a new instance $I_{n+1}$ as the set of values that the user already entered for fields in that instance:

$$C = \{(f_1, v_1), \dots, (f_m, v_m)\} \mid C \subset I_{n+1}$$

Additionally, we define the *target field $f'$* as the field that the user is about to fill in.

For example, suppose that the user is generating an instance based on the *Experiment* template and has already entered the value *male* for the field *sex* and the value *liver* for the field *tissue*. Suppose also that the user is about to enter a value for the *disease* field. In this example, the fields *sex* and *tissue* constitute the context, while *disease* is the target field:

$$C = \{(sex, male), (tissue, liver)\}$$

$$f' = disease$$

Now that we have defined the notions of template, template instance, fields, values, instance repository, and context, we can formally define our value-recommendation approach as a function $recommend$ which, given a context $C$, a target field $f'$, and an instance repository $I$, returns a ranked list of recommended values $V'$:

$$V' = recommend(C, f', I)$$

with $V' = \{v_i \in V \mid 1 \leq i \leq n\}$, such that $V$ is the set of all unique values in the instance repository and $i$ represents the position of the value in the ranking of results (e.g., $v_1$ is the top recommended value). An example of recommended values for a field named *disease* is:

$$V' = (liver\ cancer, cirrhosis, meningitis)$$

In this example, there are three recommended values for the *disease* field, with *liver cancer* as the highest ranked recommendation.

The $recommend$ function constitutes the core of our approach and it assumes that existing template instances contain hidden relationships between the values of populated fields that can be used to generate value recommendations for yet-to-be-populated fields.

For example, suppose that when the field *disease* has the value *meningitis*, the *tissue* field always has the value *brain*. Suppose now that a user is creating a new template instance and has already entered the value *meningitis* for the *disease* field. Then, the $recommend$ function should be able to suggest the value *brain* for the *tissue* field.

Of course, the relationships between field–value pairs can be far more complex than this simple example. In a real scenario, the instance repository may contain thousands of template instances with millions of different relationships among the values of the fields, many of them of little or no significance. It is therefore desirable to have a method to efficiently extract and represent all the relationships, together with some additional information indicating how reliable the relationships are.

In this work, we address this problem by using a data mining technique known as *Association Rule Mining* (ARM) (25). This method can be used to discover interesting associations among values and to represent them in the form of if–then statements known as *association rules* (or simply *rules*). We use association rule mining to extract association rules from a set of existing template instances. Then, these rules are used to generate a ranked list of values for the target field.

Our approach encompasses three steps. The first step, *rule extraction*, produces relevant association rules from an instance repository. The second step, *rule matching*, selects the best rules to generate value recommendations based on a particular context and target field. Finally, the third step, *value ranking*, generates and returns a ranked list of recommended values for the target field. Figure 1 shows a schematic representation of the approach.

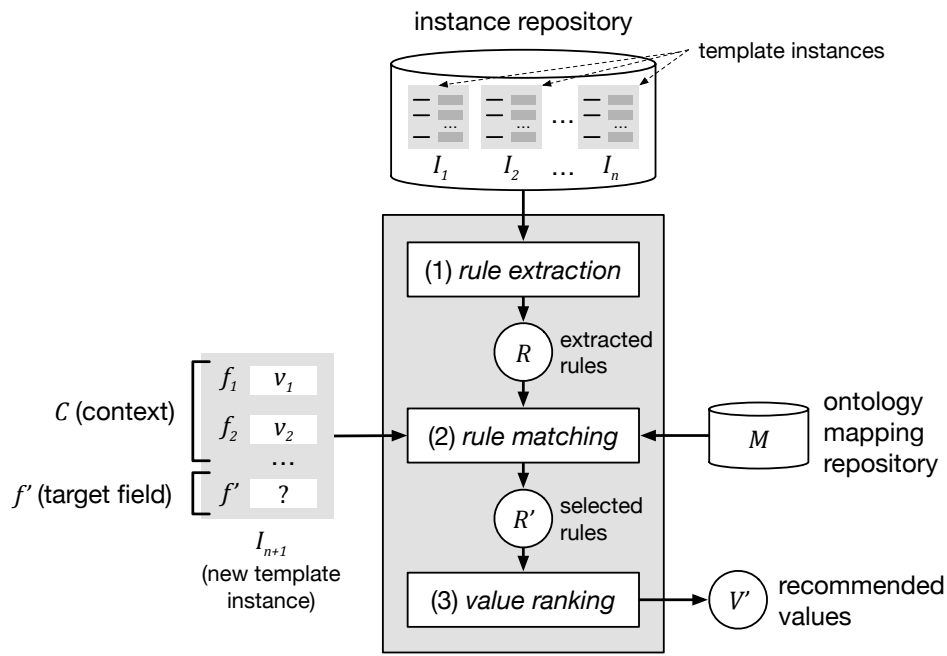

Figure 1. Schematic representation of the value-recommendation approach. Here, a user has entered values for a new template instance $I_{n+1}$. We refer to the values that the user already entered (i.e., $v_1$ and $v_2$) as the *context* ($C$). The field that the user is about to fill out is known as the *target field* ($f'$). The instance repository stores all the template instances previously created. The value-recommendation process uses the new instance plus all existing instances to generate recommendations for the target field using the following three steps: (1) Rule extraction: extract relevant association rules from the repository; (2) Rule matching: select the most appropriate rules to generate the recommended values; (3) Value ranking: rank and return the recommended values.

### 3.1.1 Rule extraction

The goal of the *rule extraction* process is to discover relevant relationships between field–value pairs in the instance repository and to represent those relationships as *association rules*. These rules will be used later to predict the value of a target field. In our approach, we define *rule extraction* as a function $rules$ which, given an instance repository $I$, applies an association rule mining algorithm to return a set of association rules $R$:

$$rules(I) = R, \text{ with } R = \{r_1, r_2, \ldots, r_n\}$$

where $r_i$ is an association rule.

An association rule can be generally defined as an implication expression of the form $X \rightarrow Y$, where $X$ and $Y$ are disjoint sets of items (26). Informally, an association rule can be read as meaning that if all items in $X$ are true then all items in $Y$ must be true. In our approach, the items are field–value pairs. $X$ corresponds to the left-hand side or *antecedent* of the rule, while $Y$ corresponds to the right-hand side or *consequent* of the rule. Since the goal is to generate value recommendations for just one target field at a time, we restrict the rule-generation process to rules that have only one attribute–value pair in the consequent. Therefore, a rule $r_i$ can be represented as:

$$(f_1, v_1) \wedge (f_2, v_2) \wedge \ldots \wedge (f_n, v_n) \rightarrow (f_j, v_j)$$

An example rule extracted from the instances shown in Table 1 appears as follows:

$$(sex = male) \wedge (disease = meningitis) \rightarrow (tissue = brain)$$

The rule indicates that there is a relationship between the fields *sex*, *disease*, and *tissue*, such that when the value of *sex* is *male* and the value of *disease* is *meningitis*, the value of *tissue* is *brain*. This rule could be used to predict the value of the *tissue* field when the user has already entered values for *sex* and *disease*.

In association rule mining, the strength of a rule is generally expressed in terms of its *support* and *confidence*. Support corresponds to the number of template instances in the repository that include all field–value pairs in the rule. For example, the support of the rule shown above is 2, because, as shown in Table 1, there are two instances in the repository ($I_1$ and $I_6$) containing the field–value pairs *sex=male*, *disease=meningitis*, and *tissue=brain*. Confidence represents how frequently the rule consequent appear in instances that contain the antecedent. It effectively measures the reliability of the inference made by the rule. The confidence of the previous rule is 1, because, when *tissue=brain*, the fields *sex* and *disease* always have the values *male* and *meningitis*, respectively. In the following rule:

$$(tissue = liver) \rightarrow (disease = liver\ cancer)$$

support is also 2, because there are two instances in the repository depicted in Table 1 that support the rule ($I_4$ and $I_5$). However, confidence is 2/3=0.67. Confidence is lower for this rule because there are three instances that contain *tissue=liver* ($I_3$, $I_4$, and $I_5$) but only two of them contain *disease=liver cancer* ($I_4$ and $I_5$).

Table 2. Example of association rules extracted from the instances shown in Table 1 together with their support and confidence. The rules are ordered first by confidence and then by support.

| Rule | Rule content | Support | Confidence |
| --- | --- | --- | --- |
| $r_1$ | $(disease = meningitis) \rightarrow (tissue = brain)$ | 3 | 1 |
| $r_2$ | $(tissue = brain) \rightarrow (disease = meningitis)$ | 3 | 1 |
| $r_3$ | $(disease = liver\ cancer) \rightarrow (tissue = liver)$ | 2 | 1 |
| $r_4$ | $(sex = male) \wedge (disease = meningitis) \rightarrow (tissue = brain)$ | 2 | 1 |
| $r_5$ | $(sex = male) \wedge (tissue = brain) \rightarrow (disease = meningitis)$ | 2 | 1 |
| $r_6$ | $(sex = female) \rightarrow (tissue = brain)$ | 1 | 1 |
| $r_7$ | $(sex = female) \rightarrow (disease = meningitis)$ | 1 | 1 |
| $r_8$ | $(disease = cirrhosis) \rightarrow (tissue = liver)$ | 1 | 1 |
| $r_9$ | $(sex = male) \wedge (disease = liver\ cancer) \rightarrow (tissue = liver)$ | 1 | 1 |
| $r_{10}$ | $(sex = male) \wedge (tissue = liver) \rightarrow (disease = liver\ cancer)$ | 1 | 1 |
| $r_{11}$ | $(sex = female) \wedge (disease = meningitis) \rightarrow (tissue = brain)$ | 1 | 1 |
| $r_{12}$ | $(sex = female) \wedge (tissue = brain) \rightarrow (disease = meningitis)$ | 1 | 1 |
| $r_{13}$ | $(tissue = brain) \rightarrow (sex = male)$ | 2 | 2 / 3 = 0.67 |
| $r_{14}$ | $(sex = male) \rightarrow (tissue = brain)$ | 2 | 2 / 3 = 0.67 |
| $r_{15}$ | $(disease = meningitis) \rightarrow (sex = male)$ | 2 | 2 / 3 = 0.67 |
| $r_{16}$ | $(sex = male) \rightarrow (disease = meningitis)$ | 2 | 2 / 3 = 0.67 |
| $r_{17}$ | $(tissue = liver) \rightarrow (disease = liver\ cancer)$ | 2 | 2 / 3 = 0.67 |
| $r_{18}$ | $(tissue = brain) \rightarrow (disease = meningitis)$ | 2 | 2 / 3 = 0.67 |

Table 2 shows all the rules extracted from the instances shown in Table 1. In our approach, all values are assumed to be categorical. Continuous attributes could potentially be managed by converting them to categorical values before applying the rule-generation algorithm.

### 3.1.2 Rule matching

The rule matching step selects the subset of association rules that can be used to generate value recommendations. This step has two stages. The first stage identifies the rules that can produce values for a target field; we refer to these rules as the *selected rules*. The second stage uses the context to rank the selected rules.

Given a set of rules $R$ and a target field $f'$, the selected rules are defined as the subset of rules $R' \subseteq R$ whose consequent matches the target field $f'$. The consequent of the selected rules contains values for the target field that are effectively candidates to generate value recommendations.

As an example, suppose that we have the rules shown previously (see Table 2) and that we are filling in a form where the target field $f'$ is $tissue$. In this case, the selected rules $R'$ are those rules for which the consequent of the rule matches the *tissue* field, that is:

$$R' = \{r_1, r_3, r_4, r_6, r_8, r_9, r_{11}, r_{14}\}$$

Suppose now that the context, that is, the field–value pairs that the user has already filled out is:

$$C = \{(disease, meningitis)\}$$

All the selected rules contain candidate values for the target field. However, not all the selected rules match the context to the same degree, and therefore not all the candidate values will be equally relevant. In order to quantify the relevance of the selected rules, our method calculates a score that measures the similarity between the antecedent of the rule and the field–value pairs that the user already filled out. That score is known as the *context-matching score*.

We define $context\_matching\_score: R' \times C \rightarrow [0,1]$ as a function that measures the degree of similarity between the antecedent of a selected rule $r' \in R'$ and the context $C$. Since both the antecedent of the rule and the context are sets of field–value pairs, we can easily calculate the similarity between them using the Jaccard Index ($J$), also known as intersection over union, which is a statistic commonly used to quantify the similarity between two sets of items:

$$context\_matching\_score(r', C) = J(antecedent(r'), C) = \frac{|antecedent(r') \cap C|}{|antecedent(r') \cup C|} \text{ with } r' \in R'$$

The most highly rated rules will be those whose antecedent best matches the values already entered by the user. A context-matching score of 1 means that the antecedent of the rule matches all the field–value pairs in the context, while a score of 0 means that there is no match. Table 3 shows an example set of context-matching scores for the selected rules in the example above.

Table 3. Selected rules and corresponding context-matching scores for the rules in Table 2 when the target field is *tissue* and with the context that the *disease* field's value is *meningitis*.

| Rule | Value | Context-matching score |
|---|---|---|
| $r_1$ | brain | 1 / 1 = 1 |
| $r_3$ | liver | 0 / 2 = 0 |
| $r_4$ | brain | 1 / 2 = 0.5 |
| $r_6$ | brain | 0 / 2 = 0 |
| $r_8$ | liver | 0 / 2 = 0 |
| $r_9$ | liver | 0 / 3 = 0 |
| $r_{11}$ | brain | 1 / 2 = 0.5 |
| $r_{14}$ | brain | 0 / 2 = 0 |

### 3.1.3   Value ranking

The final step of the value-recommendation process uses the selected rules and the associated context-matching scores to generate a ranked list of recommended values for the target field. The candidate values for the target field are extracted from the consequent of the selected rules. Then, these values are ranked according to a *recommendation score* that provides an absolute measurement of the goodness of the recommendation.

The recommendation score is based on two factors:

1. **The context-matching score**, which represents the degree of similarity between the antecedent of the rule and the context entered by the user. The most relevant rules will be those whose antecedent best matches the values already entered by the user, since those rules represent more closely the template instance that the user is creating.
2. **The rule confidence**, which reflects the proportion of consequents predicted by the rule that are correctly predicted. It is a measure of the reliability of the inference made by the rule and therefore it represents how trustworthy the recommended value is. Confidence is the primary metric for ranking rules in associative classification problems. We have also considered using the rule lift as an alternative to confidence and conducted a small experiment to compare their performance. The results show that confidence performs better than lift to rank the suggested values.[1]

Suppose that $v'$ is a value for the target field extracted from the consequent of a selected rule $r'$. Then, the recommendation score of $v'$ is a score in the interval [0,1] that is calculated as follows:

$$recommendation\_score(v') = context\_matching\_score(r', C) * conf(r')$$

where $context\_matching\_score: R' \times C \rightarrow [0,1]$ is the function that computes the context-matching score and $conf: R \rightarrow [0,1]$ is the function that returns the confidence of a particular rule.

When there is no context (i.e., the user has not yet entered any values), the recommendation score is calculated as the rule support, which serves as an indicator of the frequency of a particular value, normalized to the interval [0,1]. Values with the same recommendation score are sorted by support. Values with a recommendation score of 0 are discarded. In the case of duplicated values, we pick the value with the highest recommendation score. The approach can be optionally adapted to discard values below a specific cutoff recommendation score.

Table 4 shows the recommendation scores for the previous example. In this case, our recommendation approach would return only the value *brain* for the *tissue* field, with a recommendation score of 1. This is a useful value in this case, since *meningitis* is a disease that affects the membranes that enclose the brain and the spinal cord.

---

[1] The results of this experiment are available in our Jupyter notebook (see "Additional experiment 1" at https://goo.gl/GtK956).

Table 4. Selected rules and corresponding values for the example in Section 3.1.2. The top recommended value is *brain*. The column *Value* contains the values of the target field, extracted from the consequent of the rule. The column *Rank* contains the position of the value in the ranking of recommended values. N/A means that the value was discarded, either because its recommendation score was 0, because it was a duplicate, or both.

| Rule | Value | Context matching score | Rule confidence | Recommendation score | Rank |
|---|---|---|---|---|---|
| $r_1$ | brain | 1 | 1 | 1 * 1 = 1 | 1 |
| $r_3$ | liver | 0 | 1 | 0 * 1 = 0 | N/A |
| $r_4$ | brain | 0.5 | 1 | 0.5 * 1 = 0.5 | N/A |
| $r_6$ | brain | 0 | 1 | 0 * 1 = 0 | N/A |
| $r_8$ | liver | 0 | 1 | 0 * 1 = 0 | N/A |
| $r_9$ | liver | 0 | 1 | 0 * 1 = 0 | N/A |
| $r_{11}$ | brain | 0.5 | 1 | 0.5 * 1 = 0.5 | N/A |
| $r_{14}$ | brain | 0 | 0.67 | 0 * 0.67 = 0 | N/A |

## 3.2 Support for ontology-based values

We have explained how our approach is able to take advantage of plain text values to generate text-based value recommendations. Additionally, our value-recommendation approach has been designed to support fields whose values are represented using ontology terms and to use them to generate ontology-based value recommendations.

For example, a template could constrain a field named *cell type* to contain specific types of cells from a branch of the Cell Ontology. For ontology-based values, each value has two components—a display label (e.g., *erythrocyte*) and a unique URI identifier (e.g., http://purl.obolibrary.org/obo/CL_0000232). Since a particular value may be associated to different display labels in different applications or data sources, our rule extraction method focuses on analyzing the relationships between the term identifiers independently of the display value used. Consider a set of template instances that refer to the disease *hepatitis B* in different ways, including *hepatitis B*, *chronic hepatitis B*, *serum hepatitis*, and *hepatitis B infection*.

Suppose also that those terms have been linked to the ontology term *hepatitis B* from the Human Disease Ontology, which has the form *obo:DOID_2043*, using *obo* as the prefix for the namespace *http://purl.obolibrary.org/obo/*. In this case, our approach would use the term identifier (1) to analyze the instances and extract the appropriate rules; (2) to match a new instance to the existing rules; and (3) to generate the list of recommendations, effectively aggregating the frequencies of all synonyms of *hepatitis B*.

For example, suppose we have the following instance:

$$(tissue = liver) \land (disease = serum\ hepatitis)$$

This instance can be represented using ontology terms as follows:

$$(obo: UBERON\_0000479 = obo: UBERON\_0002107) \land (obo: DOID\_4 = obo: DOID\_2043)$$

Similarly, the rule:

$$(disease = serum\ hepatitis) \rightarrow (tissue = liver)$$

can be represented using ontology terms as:

$$(obo: DOID\_4 = obo: DOID\_2043) \rightarrow (obo: UBERON\_0000479 = obo: UBERON\_0002107)$$

Finally, given that the content of the rules is encoded using ontology terms, the recommended values are represented using ontology terms as well, which can be visually presented to the user using the preferred label for the ontology term defined in the source ontology. For example, *hepatitis B* is the preferred label for *obo:DOID_2043* in the Uber Anatomy Ontology.

### 3.3 Support for cross-template recommendations

One of the most innovative aspects of our approach is that it has been designed to take advantage of semantic mappings between ontology terms to generate recommendations based on metadata, not only from one template, but also from multiple, structurally different templates.

We have described how our framework takes advantage of the hidden relationships between fields and their values in existing template instances to generate rules that offer value recommendations for fields in new template instances. In our previous explanations, we have assumed that the repository was pre-populated with multiple instances of a particular template and described the value-recommendation process when entering new instances for that template.

However, in general, a repository may contain multiple instances not only from one template but also from multiple templates. Those templates can have different field names and field values, but they may also store information about the same concepts. For instance, suppose we have an *Assay* template with a field *cell type* that indicates the type of cell under analysis. Suppose also that we have an *Experiment* template with a field *source cell* that captures the same kind of information. Ideally, our method should be able to determine that these fields are equivalent and to analyze the values of the two fields as a single set to generate recommendations.

Our method uses ontologies to determine the correspondences between the same fields occurring in different templates with different names (e.g., *cell type* and *source cell*) and between the same values used with different names (e.g., *erythrocyte* and *red blood cell*). When working with plain text instances, the field–value pairs of the instance that the user is creating are matched to the association rules by comparing the display labels for fields and values. However, the full potential of our method is reached when working with ontology-based instances. In that case, the matching is done at a semantic level, by comparing the ontology term URIs linked to the fields and values.

For example, suppose that the fields *cell type* and *source cell* are annotated with the term *cell type* from the Experimental Factor Ontology (EFO) (*efo:EFO_0000324*). Given that both fields are linked to the same term, our method will determine that they are equivalent, and it will treat them as if they were the same type of field when performing the rule extraction and rule matching steps.

It is common to find ontology terms with different URIs in different ontologies referring to the same underlying real-world concept. For example, the term *tissue* has different URIs in the National Cancer Institute Thesaurus (NCIT) (*ncit:C12801*) and in the Uber Anatomy Ontology (UBERON) (*obo:UBERON_0000479*). As a consequence, equivalent *tissue* fields from different templates may be linked to ontology terms with the same meaning, but with different URIs. To deal with this scenario, our framework includes a component that stores correspondences or mappings between equivalent terms in different ontologies. This component is called the ontology mapping repository $M$ (see Figure 1). Given $t_i$ an ontology term URI, we define $equivalent\_terms(t_i, M)$ as the function that returns the set of all ontology term identifiers in the mapping repository that are equivalent to $t_i$: $equivalent\_terms(t_i, M) = \{t_1, t_2, ..., t_n\}$

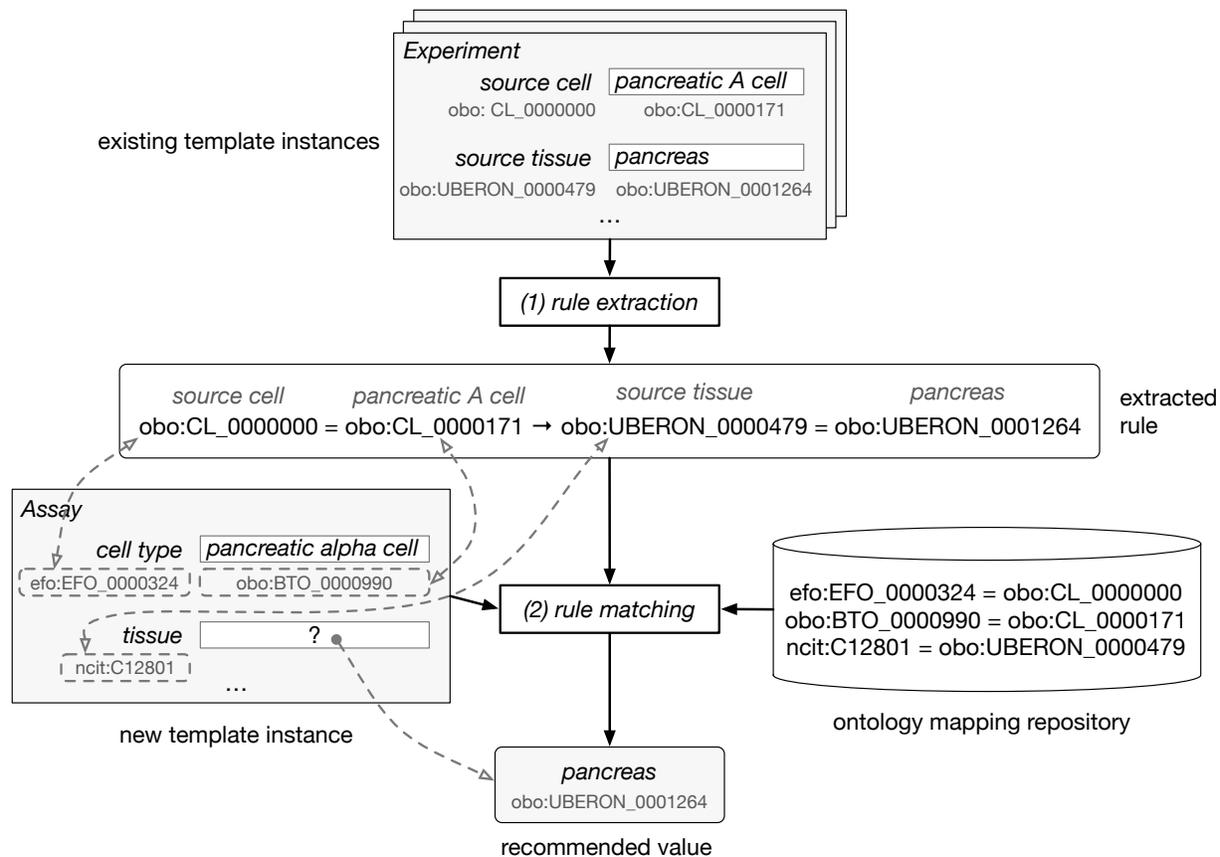

Figure 2. Example of rule extraction and rule matching process for ontology-based values. From top to bottom, the figure shows the instances of the *Experiment* template and an example of rule extracted from them. The user is creating a new instance of the *Assay* template. This instance is matched to the rule using the equivalences between ontology terms stored in the ontology mapping repository. Finally, the recommended value for the *tissue* field is *pancreas*.

Figure 2 shows how our framework uses ontology terms and a repository of ontology mappings to generate value recommendations. In this example, the user is filling out an *Assay* template with two fields: *cell type*, annotated with *efo:EFO_0000324*; and *tissue*, annotated with *ncit:C12801*. The user already has entered the value *pancreatic alpha cell* (*obo:0000990*) for the field *cell type* and is about to enter a value for the *tissue* field. Our recommendation framework will generate a list of recommended values for the *tissue* field.

The instance repository contains some instances of an *Experiment* template, which has two fields: *source cell* (*obo:CL_0000000*) and *source tissue* (*obo:UBERON_0000479*). Although these two fields have the same meaning as the fields in the *Assay* template, the field names in the *Experiment* template are different, and the ontology terms linked to them are different as well.

Our approach uses the ontology mapping repository to determine that the *cell type* and the *source cell* fields are equivalent, and that the *tissue* and *source tissue* fields are also equivalent. These correspondences make it possible to match the fields and values from the new instance of the *Assay* template to the rule extracted from the instances of the *Experiment* template and, finally, to recommend the value *pancreas* (*obo:UBERON_0001264*) for the *tissue* field in the new instance.

## 3.4  Implementation

We integrated our value-recommendation approach into a metadata collection and management platform called the CEDAR Workbench (27), which was developed by the Center for Expanded Data Annotation and Retrieval (CEDAR) (5). The CEDAR Workbench is a Web-based system comprising a set of highly-interactive tools to help create, manage, and submit biomedical metadata for use in online data repositories. The ultimate goal is to improve the metadata acquisition process by helping users enter their metadata rapidly and accurately.

CEDAR provides technology to allow scientists to create and edit metadata templates for characterizing the metadata for different types of experiments. Investigators then fill out those templates to create rich, high-quality instances that annotate the corresponding datasets. Two key tools called the Template Designer and the Metadata Editor (27,28) provide this functionality. The Template Designer allows users to build metadata templates interactively in much the same way that they would create online survey forms. Using live lookup to the NCBO BioPortal ontology repository (29), the Template Designer allows template authors to find terms in ontologies to annotate their templates, and to constrain the values of template fields to specific ontology terms (Figure 3) (30).

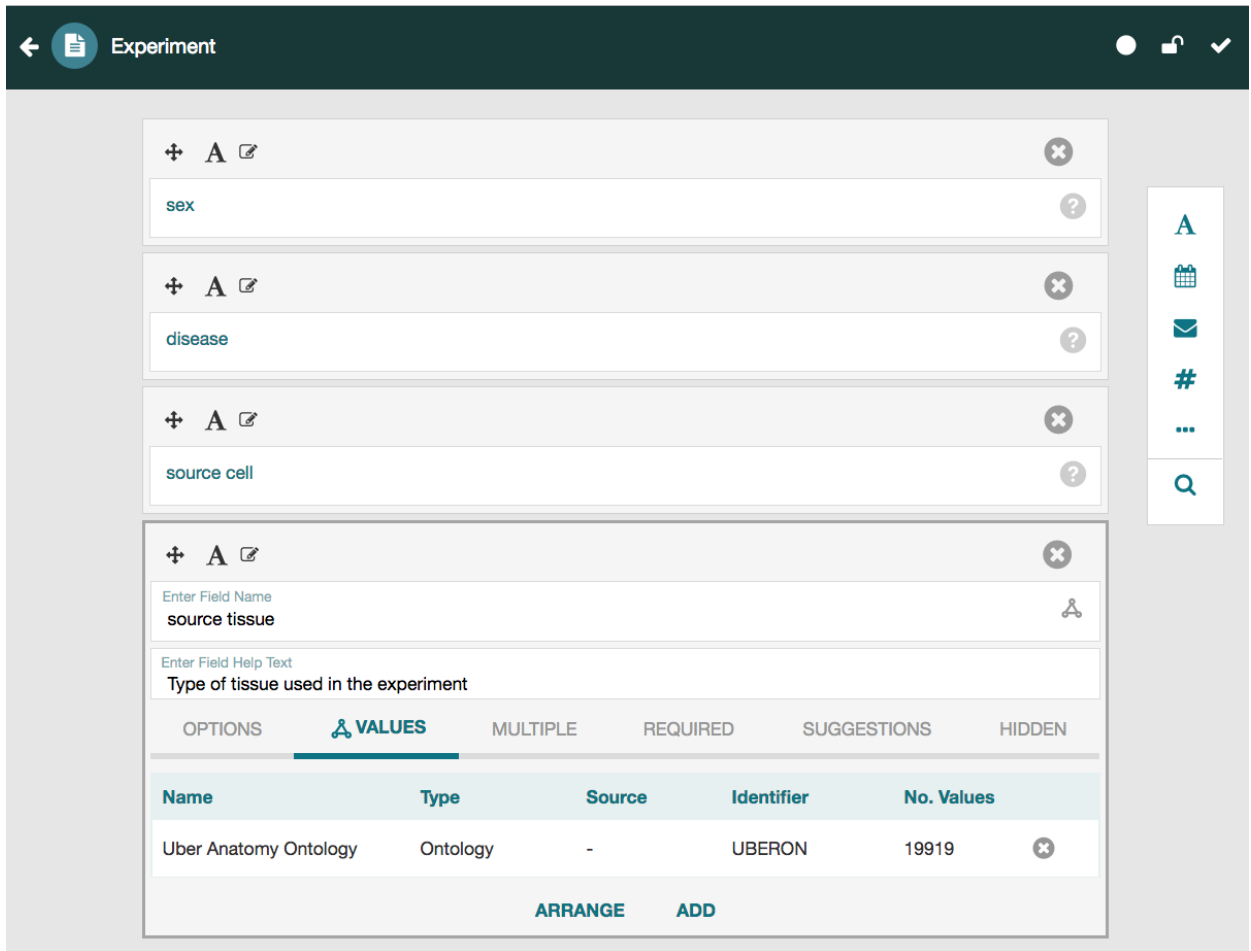

Figure 3. Screenshot of the Template Designer showing the creation of an *Experiment* template with five fields: *sex*, *disease*, *source cell*, and *source tissue*. The user can interactively add fields of predefined types (text, date, email, numeric, etc.) to the template and specify additional configuration options for each field. When appropriate, template fields are linked to value sets, ontologies, or branches of ontologies stored in the BioPortal repository, standardizing potential values of those fields. Here, the user has specified that values for the field *source tissue* should come from the Uber Anatomy Ontology.

The Metadata Editor (Figure 4) uses these templates to automatically generate a forms-based acquisition interface for entering metadata. It also uses live lookup to BioPortal to provide selection lists for metadata authors filling out fields. The values in these lists are generated using the constraints specified for fields by the associated template.

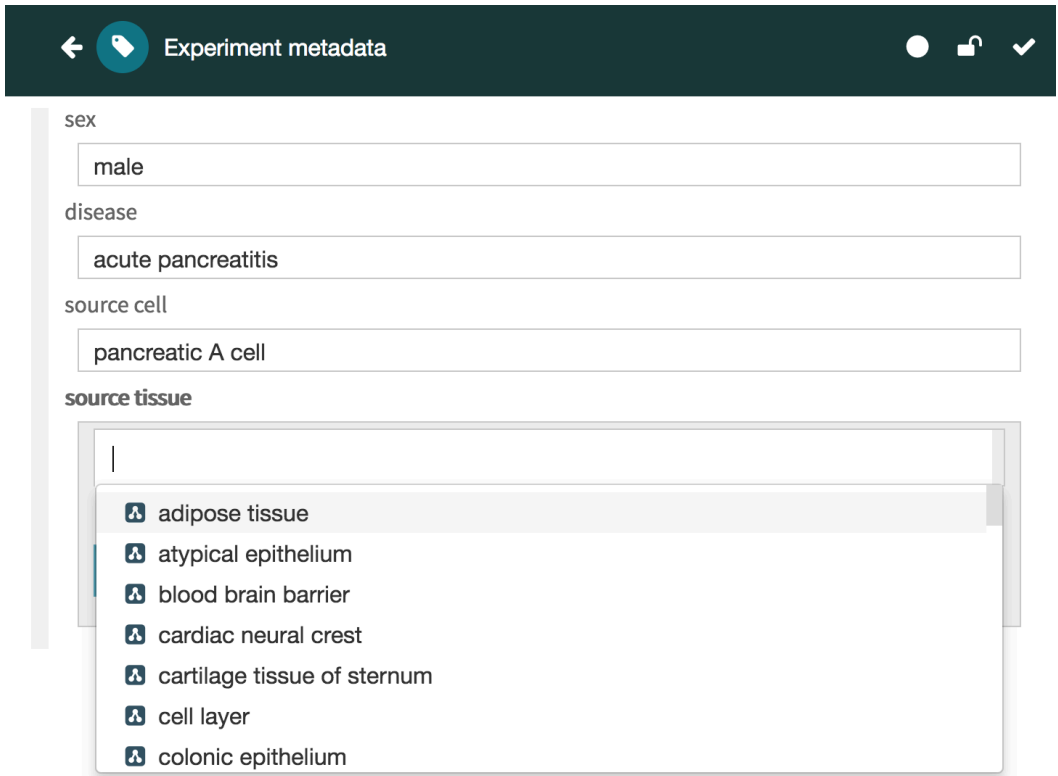

Figure 4. Screenshot of the Metadata Editor for the *Experiment* template showing a list of valid values for the *source tissue* field, which were constrained to the Uber Anatomy Ontology (see Figure 3).

The CEDAR Workbench aims to ensure that users can quickly create high-quality, semantically rich metadata and submit these metadata to external repositories. CEDAR commits to the FAIR principles (6) in terms of standards, protocols, and best practices. Metadata created using the CEDAR Workbench are represented using standard formats and stored in a searchable, centralized repository. Users can search for metadata through either a Web-based tool or a REST API, link their metadata to terms from formal ontologies and controlled terminologies, and enrich them with a variety of additional attributes, including provenance information. Regarding FAIRness of the experimental datasets, the CEDAR Workbench can help to improve adherence of the experimental datasets to the FAIR principles by enhancing datasets findability, interoperability, and reusability. However, ensuring data accessibility is entirely at the discretion of the data owner, or the owner of the repository where the data are stored.

We implemented our value-recommendation approach as a CEDAR microservice that is used by the Metadata Editor to help users create metadata. The service is known as CEDAR's Value Recommender service.

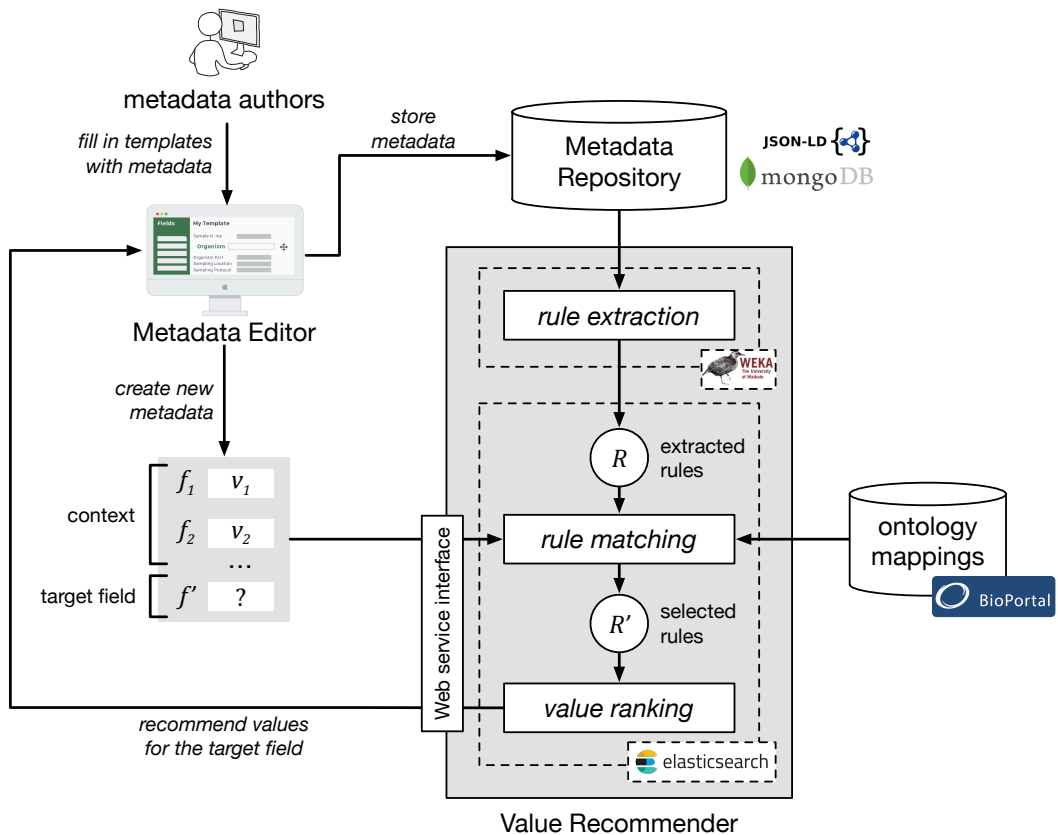

Figure 5. Architecture and workflow of CEDAR's Value Recommender service. Metadata authors use the Metadata Editor to create metadata. Entered metadata are stored in the MongoDB-based Metadata Repository. The association rules are extracted from existing metadata using WEKA's software and are stored in Elasticsearch. The rule matching step uses BioPortal's ontology mappings to determine the correspondences between the fields and values in the template that the user is filling out and the extracted rules. Finally, the ranked list of recommended values is returned to the Metadata Editor and presented to the user.

Figure 5 shows the architecture and workflow of CEDAR's Value Recommender service and the main integration points with CEDAR's Metadata Editor and the BioPortal ontology repository. Metadata authors use the Metadata Editor to generate metadata instances based on templates. Entered metadata are stored in CEDAR's Metadata Repository as a JSON document using an open, standards-based model (31). This model represents templates and metadata using JSON-LD constructs (32), making it possible to restrict the types and values of fields to terms from ontologies. JSON-LD is an RDF serialization, so CEDAR can use off-the-shelf tools to export metadata in a variety of RDF serialization formats (e.g., Turtle, RDF/XML). The Metadata Repository is implemented using the MongoDB database, which is a NoSQL database based on JSON that provides fast and scalable storage.

```json
{
  "antecedent": [
    {
      "fieldLabel": "source cell",
      "fieldType": "http://purl.obolibrary.org/obo/CL_0000000",
      "fieldTypeMappings": ["http://www.ebi.ac.uk/efo/EFO_0000324", ...],
      "fieldValueLabel": "pancreatic A cell",
      "fieldValueType": "http://purl.obolibrary.org/obo/CL_0000171",
      "fieldValueMappings": ["http://purl.obolibrary.org/obo/BTO_0000990", ...]
    }
  ],
  "consequent": [
    {
      "fieldLabel": "source tissue",
      "fieldType": "http://purl.obolibrary.org/obo/UBERON_0000479",
      "fieldTypeMappings": [
        "http://ncicb.nci.nih.gov/xml/owl/EVS/Thesaurus.owl#C12801", ...],
      "fieldValueLabel": "pancreas",
      "fieldValueType": "http://purl.obolibrary.org/obo/UBERON_0001264",
      "fieldValueMappings": ["http://purl.obolibrary.org/obo/BTO_0000998", ...],
    }
  ],
  "support": 213,
  "confidence": 0.6923076923076923,
  "templateId": "https://repo.metadatacenter.orgx/templates/2e4f3c1bd179"
}
```

Figure 6. Example of a generated association rule stored in Elasticsearch. The rule is represented as a JSON document that contains the field–value pairs in the left-hand side of the rule (*antecedent*), the field–value pair in the right-hand side of the rule (*consequent*), the rule support and confidence, and the identifier of the source template (*templateId*). For the field–value pairs in the antecedent and consequent, we store the field name (*fieldLabel*), the URI of the ontology term linked to the field (*fieldType*), a list of equivalent ontology terms (*fieldTypeMappings*), the field value in textual format (*fieldValueLabel*), the ontology term linked to the field value (*fieldValueType*), and a list of equivalent ontology terms for it (*fieldValueMappings*).

CEDAR's Value Recommender has been implemented in Java and uses the Dropwizard framework (https://www.dropwizard.io). It provides a REST-based API that is used by the Metadata Editor and can also be used directly by third-party applications. The rule extraction process is performed using the Java API of the WEKA data mining software (https://www.cs.waikato.ac.nz/ml/weka). All metadata for a particular template are transformed to WEKA's ARFF format (https://www.cs.waikato.ac.nz/ml/weka/arff.html). The association rules are extracted using the Apriori algorithm (33), which is commonly used in association rule mining.

The extracted rules are stored using the Elasticsearch engine (https://www.elastic.co/products/elasticsearch) in JSON. We defined a custom JSON format designed to represent rule antecedents and consequents as field–value pairs, together with relevant rule metrics. Figure 6 contains an example of how an association rule is stored in Elasticsearch.

For plain text metadata, our rule model stores the field name and the textual value (i.e., *fieldLabel* and *fieldValueLabel*). Additionally, for ontology-based metadata, it stores the ontology terms linked to the field name and to the field value, as well as an array of equivalent ontology terms extracted from BioPortal's ontology mapping repository, which is accessed via the BioPortal API (http://data.bioontology.org/documentation). The rule support and confidence are also stored, as well as the identifier of the template from which the metadata are derived.

The rule matching step takes advantage of the search flexibility offered by the Elasticsearch engine, particularly, the MUST and SHOULD filters:

1. To perform a query that selects the rules whose consequent matches the target field. This is done by means of an Elasticsearch MUST filter.
2. To calculate a matching score for each rule that reflects the degree of similarity between the rule antecedent and the context. This action is done using a SHOULD filter.

Finally, the values are ranked according to their recommendation score and returned in JSON format via the REST API to the Metadata Editor, which presents the recommended values to the user in a drop-down list, followed by any other valid values for the target field (Figure 7). Each recommended value is accompanied by its recommendation score presented as a percentage for better readability (e.g., 0.28 is presented as 28%). When returning ontology-based recommendations, the values are presented using a user-friendly label for the ontology term defined in its source ontology (e.g., *pancreas* is the preferred label for the term *obo:UBERON_0000479* in the Uber Anatomy Ontology).

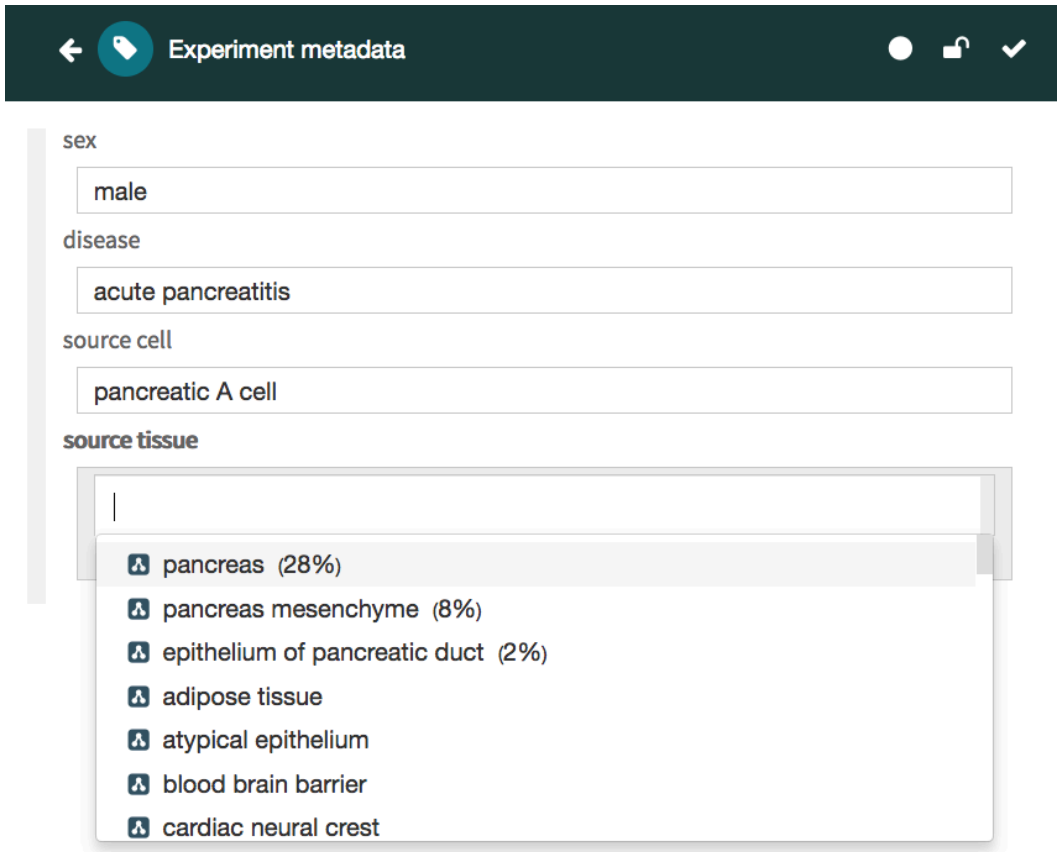

Figure 7. Screenshot of the CEDAR Metadata Editor showing recommended values for a particular target field. In this case, the editor shows three suggested values for the *source tissue* field ranked in order of likelihood: *pancreas*, *pancreas mesenchyme*, and *epithelium of pancreatic duct*, followed by other valid values for the target field. Ontology-based terms are indicated with an ontology icon. The recommendation score for each suggested value is presented as a percentage.

## 4 Evaluation

We evaluated CEDAR's Value Recommender by measuring its accuracy when generating suggestions for (1) single-template recommendations, where the recommendations for the template that the user is filling out are based on metadata created using the same template; and (2) cross-template recommendations, where the recommendations are based on metadata created using a different template. For each of these two scenarios, we analyzed the behavior of our system for two different kinds of metadata: (a) text-based, where the field values are free text; and (b) ontology-based, where the values are ontology terms. We designed a total of 8 experiments to cover all combinations of recommendation scenario (single-template or cross-template) and metadata type (text-based or ontology-based).

We used a subset of metadata from two public biomedical databases: NCBI BioSample (34) and EBI BioSamples (35). We applied a "train and test" approach that is commonly used to evaluate data mining models. We split each dataset into a training set and a test set. First, the training set was used to discover the hidden relationships between metadata fields and to represent them as association rules. The association rules were then used to generate recommendations for the values of the fields in the test set. Because the test set already contains values for the target field, it is straightforward to determine whether the system's suggestions are correct.

The NCBI BioSample and EBI BioSamples databases contain descriptive metadata for diverse types of biological samples from multiple species. These metadata are encoded as field–value pairs. Typical examples of metadata for a biological sample include the source organism (e.g., an *organism* field with a value of *Homo sapiens*), the cell type (e.g., a *cell_type* field with a value of *monocyte*), and the source tissue (e.g., a *tissue* field with a value of *skin*). These two databases are appropriate for our evaluation because they contain metadata about the same domain, they are publicly available, they are widely known and used in the biomedical community, and they contain a large amount of rich metadata about biomedical samples.

We constructed an evaluation pipeline to drive the analysis workflow (Figure 8). This pipeline consists of 7 sequential steps: (1) content download from NCBI BioSample and EBI BioSamples databases; (2) template design for each of those databases and generation of the corresponding template instances; (3) linkage of template instance field names and values to ontology terms; (4) dataset splitting into training and test sets; (5) generation of rules to drive the recommendations from the training set; (6) accuracy measurement using the test set; and (7) results analysis. These steps are now described in more detail.

## 4.1 Step 1: Datasets download

We downloaded the full content of the NCBI BioSample repository in XML format using NCBI's FTP service (https://ftp.ncbi.nih.gov/biosample). The resulting XML file contained metadata on 7.8M samples from multiple organisms. In the case of the EBI BioSamples repository, we downloaded a total of 4.1M samples in JSON format using EBI's REST API

(https://www.ebi.ac.uk/biosamples/docs/references/api/overview). Both datasets were downloaded on March 9, 2018.

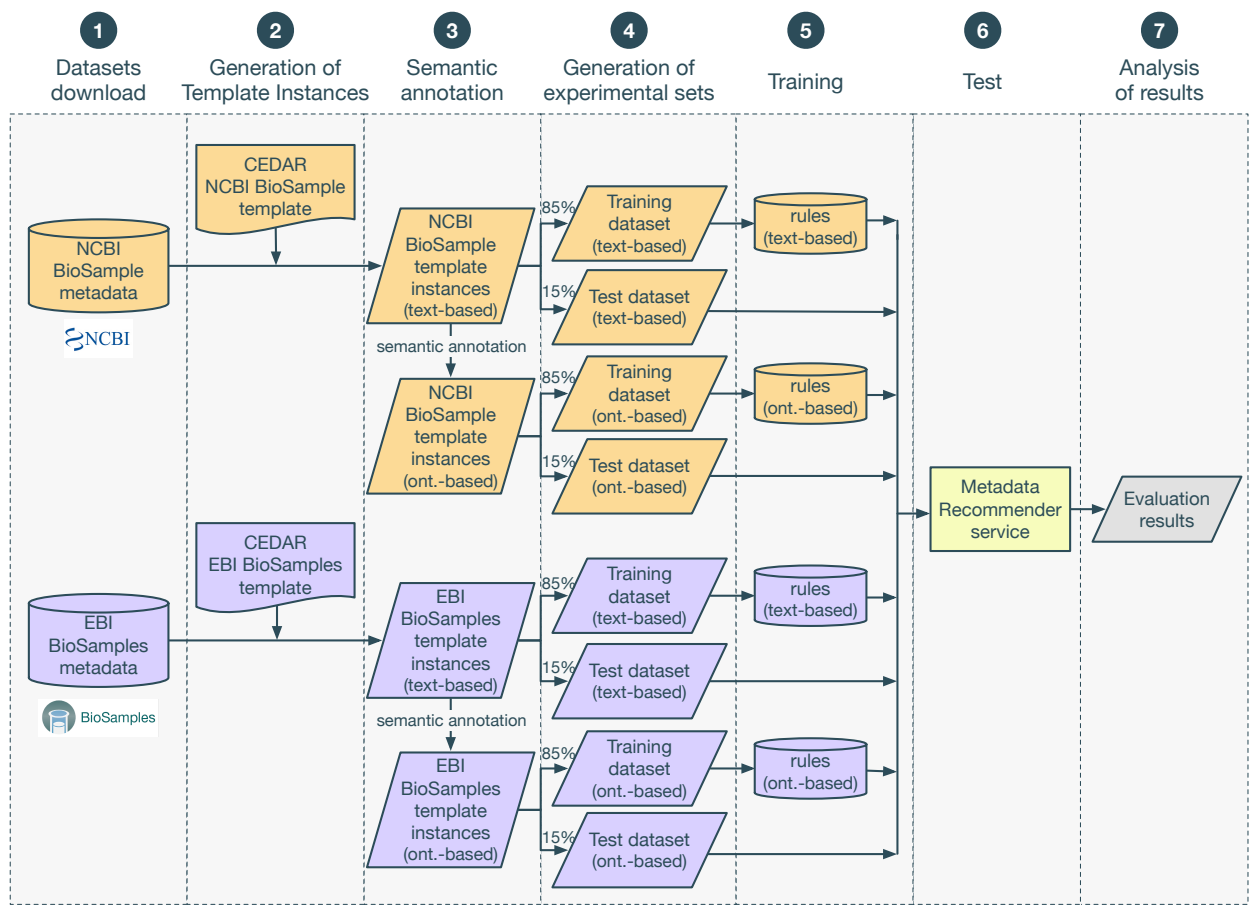

Figure 8. Steps in the evaluation pipeline. (1) NCBI BioSample and EBI BioSamples dataset download; (2) Template design for NCBI BioSample and EBI BioSamples, and instance population with metadata from the downloaded datasets; (3) Semantic annotation of text-based instances using terms from biomedical ontologies to generate ontology-based instances; (4) Partitioning into training and test datasets; (5) Rule extraction from the training dataset; (6) Generation of recommendations for the test instances, for the fields *sex*, *organism part*, *cell line*, *cell type*, *disease*, and *ethnicity*; (7) Comparison of the recommendations obtained using CEDAR's Value Recommender with the recommendations obtained using the baseline method.

## 4.2  Step 2: Generation of template instances

Both repositories contain a very large number of samples from multiple species. *Homo sapiens* is one of the most common organisms, with 4.6M samples in NCBI BioSamples (59%) and 1.4M samples in EBI BioSamples (34%). We decided to focus our evaluation on human samples because the number of samples available is large enough to design a robust evaluation. There is also a strong overlap between the metadata attributes of these samples in both repositories, which is key to produce meaningful cross-database recommendations. For each repository, we used the CEDAR Workbench to design a metadata template targeted to human samples. The NCBI BioSample repository defines several packages, which specify the list of metadata fields that should be used to describe a particular sample type. We created a metadata template that corresponds to the specification of the BioSample Human package v1.0 (https://submit.ncbi.nlm.nih.gov/biosample/template/?package=Human.1.0&action=definition), which is designed to capture metadata from studies involving human subjects. It includes a total of 26 fields. The EBI BioSamples repository does not have an equivalent formal specification. Instead, we defined a metadata template containing 14 fields with general metadata about biological samples and some additional fields that capture specific characteristics of human samples. Table 5 lists the fields of these two CEDAR templates.[2] Some fields capture the same kind of information using different field names (e.g., *cell_line* and *cellLine*). They also contain fields that store different types of values. Examples include numeric fields (e.g., *age*), free text fields (e.g., *sample_title*), identifier fields (e.g., *biosample_accession*), and fields with values limited to a finite set of choices (e.g., *ethnicity*).

Table 5. Names of the two templates used to evaluate CEDAR's Value Recommender and names of the fields in each template.

| Template name | Field names |
| --- | --- |
| NCBI BioSample - Human Package 1.0 | *biosample_accession, sample_name, sample_title, bioproject_accession, organism, isolate, age, biomaterial_provider, sex, tissue, cell_line, cell_subtype, cell_type, culture_collection, dev_stage, disease, disease_stage, ethnicity, health_state, karyotype, phenotype, population, race, sample_type, treatment, description* |
| EBI BioSamples | *accession, name, releaseDate, updateDate, organization, contact, organism, age, sex, organismPart, cellLine, cellType, diseaseState, ethnicity* |

Even though these two CEDAR templates constitute a realistic representation of the metadata about human samples usually submitted to public databases, not all their fields are equally relevant to our analysis. For example, fields that contain identifiers (e.g., *biosample_accession*) cannot be used as a source of recommendations. Similarly, fields that are present in only one of the templates (e.g., *isolate*, *karyotype*) cannot be used to generate cross-template recommendations. We focused our analysis on the subset of fields that meet two key requirements: (1) they are present in both templates and, therefore, can be used to evaluate cross-template recommendations; and (2) they contain categorical values, that is, they represent information about discrete characteristics. We selected 6 fields that met these criteria. These fields are: *sex*, *organism part*, *cell line*, *cell type*, *disease*, and *ethnicity* (Table 6).

---

[2] The two templates are publicly available on the CEDAR Workbench. NCBI BioSample template: https://tinyurl.com/ybqcatsf. EBI BioSamples template: https://tinyurl.com/y96z975d.

Table 6. Fields from the NCBI BioSample and EBI BioSamples templates selected for our evaluation. The table provides a short description for each field, as well as the field names used to refer to in both in CEDAR's NCBI BioSample template and CEDAR's EBI BioSamples template.

| Field name | Description | Field name in NCBI BioSample template | Field name in EBI BioSamples template |
| --- | --- | --- | --- |
| *sex* | Physical sex of sampled organism | *sex* | *sex* |
| *organism part* | Part of the organism's anatomy or substance arising from an organism from which the biomaterial was derived | *tissue* | *organismPart* |
| *cell line* | Name of the cell line from which the sample was extracted | *cell_line* | *cellLine* |
| *cell type* | Type of cell from which the sample was extracted | *cell_type* | *cellType* |
| *disease* | Disease for which the sample was obtained | *disease* | *diseaseState* |
| *ethnicity* | Ethnicity of the subject | *ethnicity* | *ethnicity* |

Metadata in the NCBI BioSample and EBI BioSamples databases are sparse and many samples contain only one or two non-empty values. To limit the size of the evaluation while still being able to generate meaningful context-based recommendations, we only used the samples with non-empty values for at least 3 of the 6 selected fields. As a result, we obtained 157,653 samples from NCBI BioSample and 135,187 samples from EBI BioSamples. We randomly discarded 22,466 NCBI BioSample samples and obtained two datasets with exactly the same number of samples. We transformed these samples into CEDAR template instances conforming to CEDAR's JSON-based model and finally obtained 135,187 instances of CEDAR's NCBI BioSample template and 135,187 instances of CEDAR's EBI BioSamples template.

## 4.3 Step 3: Semantic annotation

Our evaluation studied the recommendations provided by CEDAR's Value Recommender for two different kinds of metadata fields: text-based and ontology-based. We wanted to determine to what extent our system was able to take advantage of the standardization capabilities of ontologies to generate more useful recommendations. As illustrated in Figure 8, we started the semantic annotation step with two datasets: text-based NCBI BioSample instances and text-based EBI BioSamples instances. Our goal was to produce two additional datasets: ontology-based NCBI BioSample instances and ontology-based EBI-BioSamples instances.

We created a copy of the template instances generated in the previous step and linked both their fields names and values to ontology terms. This process is typically referred to as *semantic annotation* (or simply *annotation*) and can be defined more generally as the process of finding a correspondence or relationship between a term in plain text and an ontology term that specifies the semantics of that term. For example, a possible result of annotating the plain text value *liver* could be the ontology term *liver* from the Uber Anatomy Ontology, which is identified by the URI *obo:UBERON_0002107*.

We used the NCBO Annotator (36) via the BioPortal API (http://data.bioontology.org/documentation) to automatically annotate a total of 270,374 template instances (135,187 instances for each template). Table 7 summarizes the semantic annotation results. For each of the relevant fields, the table shows the number of unique text-based values and the number of ontology terms resulting from annotating them. As expected, the total number of ontology terms obtained for each set of template instances (12,177 and 4,233) is less than the number of text-based values (23,086 and 9,730), because a single ontology term can be represented using different text strings. For example, we observed that the concept *male* was represented in plain text using values such as *male*, *Male*, *M*, and *XY*. The annotation ratio represents the relation between the number of plain text values and the number of ontology terms obtained after annotating them. For higher field annotation ratios, fewer ontology terms were needed to cover all its values.

We also found multiple values that could not be mapped to ontology terms, either because they were invalid values or because the NCBO Annotator was not able to find a suitable ontology term for them. Examples of some invalid values found for the *ethnicity* field are *C?, U, not sure* *if she is hispanic of latino, Father is half iranian*, and *usa*. In total, 13% of plain text values from NCBI BioSample and 14% from EBI BioSamples were not annotated with ontology terms and were therefore ignored.

Table 7. Summary of annotation results for the NCBI BioSample and EBI BioSamples template instances. For each field, the table shows the number of different textual values and the number of different ontology terms obtained after annotating those values. The annotation ratio represents the relation between the number of plain text values and the number ontology terms.

| Field | NCBI BioSample instances | | | EBI BioSamples instances | | |
|---|---|---|---|---|---|---|
| | No. unique values | | Annotation ratio | No. unique values | | Annotation ratio |
| | Text | Ont. terms | | Text | Ont. terms | |
| sex | 41 | 18 | 2.28 | 29 | 12 | 2.42 |
| organism part | 2,098 | 646 | 3.25 | 1,759 | 610 | 2.88 |
| cell line | 16,697 | 9,933 | 1.68 | 3,451 | 1,936 | 1.78 |
| cell type | 1,464 | 521 | 2.81 | 1,456 | 526 | 2.77 |
| disease | 2,144 | 887 | 2.42 | 2,399 | 984 | 2.44 |
| ethnicity | 642 | 172 | 3.73 | 636 | 165 | 3.85 |
| **Total** | 23,086 | 12,177 | 1.89 | 9,730 | 4,233 | 2.30 |

In some cases, the NCBO Annotator returned different URIs for a plain text value (e.g., *obo:EHDA_9373* and *ncit:C46112* for the value *male*). These multiple matches are to be expected, since some biomedical terms are defined in multiple ontologies. As explained in Section 3.3, our recommendation approach is designed to deal with this case, since it is able to find the correspondences between metadata from different templates even when different ontology terms are used to represent the same concepts.

### 4.4 Step 4: Generation of experimental data sets

We partitioned each of the four datasets from the previous step into two datasets, with 85% of the data for training and the remaining 15% for testing. We ensured that the training and test sets were disjoint. We designed a total of 8 experiments to cover all combinations of recommendation scenario (single-template or cross-template) and metadata type (text-based or ontology-based) (see Table 8). For each experiment, the table shows the recommendation scenario and the type of metadata used, as well as the source databases of the training and test sets used, and the number of instances in the training and test sets. Note that for single-template recommendations (experiments 1-4), we used datasets from the same source database for training and testing. However, for cross-template recommendations (experiments 5-8), we used one dataset from one source to train and a different source to test. All the experiments were conducted on a MacBook Pro with a 3-GHz Intel Core i7 processor and 16 GB DDR3 RAM.

Table 8. Details of the 8 experiments conducted to evaluate CEDAR's Value Recommender. The table includes the recommendation scenario addressed by the experiment (single-template or cross-template), the type of metadata used (text-based or ontology-based), the source databases of the training and test sets used (NCBI BioSample or EBI BioSamples), and the number of instances (size) of the training and test sets.

| Experiment | Recommendation scenario | Type of metadata | Training set DB (size each = 114,909) | Test set DB (size each = 20,278) |
|---|---|---|---|---|
| 1 | Single-template | Text-based | NCBI | NCBI |
| 2 | Single-template | Ontology-based | NCBI | NCBI |
| 3 | Single-template | Text-based | EBI | EBI |
| 4 | Single-template | Ontology-based | EBI | EBI |
| 5 | Cross-template | Text-based | NCBI | EBI |
| 6 | Cross-template | Ontology-based | NCBI | EBI |
| 7 | Cross-template | Text-based | EBI | NCBI |
| 8 | Cross-template | Ontology-based | EBI | NCBI |

## 4.5 Step 5: Training

The training step consisted in executing the rule extraction process for the training sets to discover the hidden relationships between metadata fields and to represent them as association rules. The rules were extracted using the Apriori algorithm via the WEKA Java library with a minimum support of 5 instances and a confidence of 0.3. A given association rule can have from one or more items in the consequent. Since our framework generates recommendations for one field at a time, we filtered the resulting rules to keep only the rules with one item in the consequent. The final set of rules were indexed using Elasticsearch, following the format described earlier (see Section 3.4).

Table 9 summarizes the number of rules extracted and the execution times for each training set. The table shows, for example, that the number of rules generated for ontology-based instances is considerably lower than it is for text-based instances. For example, the number of rules generated for the NCBI BioSample database using ontology-based instances was 18,223, which constitutes 35% of the number of rules generated from text-based instances. This improvement in precision is caused by the reduction in the number of ontology-based values with respect to text-based values during the semantic annotation process. It can be also observed that the amount of time needed to generate rules from ontology-based instances was substantially lower than that for text-based instances.

Table 9. Number of rules generated by the training process and execution time (in wall-clock time) for each training set. The number of rules obtained after filtering are the rules with only one item in the consequent part of the rule.

| Training set DB | Type of metadata | No. rules generated | No. rules after filtering | Execution time (sec) |
|---|---|---|---|---|
| NCBI | Text-based | 52,192 | 30,295 | 5,682 |
| EBI | Text-based | 36,915 | 24,983 | 4,079 |
| NCBI | Ontology-based | 18,223 | 12,400 | 1,293 |
| EBI | Ontology-based | 16,838 | 11,932 | 1,087 |

## 4.6 Step 6: Testing

In the previous step, we trained CEDAR's Value Recommender by extracting association rules from the training set and incorporating them into the system. In this step, we measured how well the system uses those rules to predict the values that are actually found in our test set.

As a baseline, we used the majority class classifier, which suggests the most frequent value for each target field in the training dataset. For example, for text-based values, *male* is the value with the most occurrences for *sex* in the training data. Therefore, the baseline method always suggests *male* for that field. This baseline is widely used in the evaluation of recommendation systems because it is an intuitive and easy-to-implement way of obtaining reference results that set the minimal expected performance.

We assessed the performance of both our rule-based approach and the baseline method using the reciprocal rank (RR) statistic, which is commonly used for evaluating processes that return a ranked list of results. The RR is calculated as the inverse of the rank of the correct answer. For example, suppose that the correct value for the *disease* field is *prostate cancer*. The RR would be 1/1=1 if the system returns *prostate cancer* as the first recommended value, 1/2=0.5 if it is in the second place, 1/3=0.33 in the third place, and so on. Finally, the mean reciprocal rank (MRR) can be calculated as the mean of all the reciprocal ranks obtained.

One of the main features of our value-recommendation framework is that it takes into account the contextual information provided by the user (that is, the values already entered for the template that the user is filling out). We wanted to analyze how different amounts of contextual information (no context, one populated field, two populated fields, etc.) affect the accuracy of the recommendations. Therefore, for each blank field, we generated recommendations using a different number of populated fields, and we compared the results obtained.

For example, suppose a template instance has the following values for the six fields used in our evaluation: *sex=male, organism part=prostate, cell line=PC-3, cell type=prostate cell, disease=prostate cancer, ethnicity=Caucasian*, and that the target field is *disease*. We first generated value recommendations without any contextual information. Then, we generated suggestions based on the value of just one populated field (e.g., *cell line=PC-3*), trying all five remaining fields. Then we generated suggestions based on two populated fields for all possible pairs of fields (e.g., *cell line=PC-3, sex=male*), and so on, until we used five populated fields as context.

The number of executions for a particular number of populated fields is given by the binomial coefficient $C(n,r)$, which represents the number of combinations of *n* items taking *r* at a time. In our example, the total number of executions would be calculated as:

$$C(5,0) + C(5,1) + C(5,2) + C(5,3) + C(5,4) + C(5,5) = 1 + 5 + 10 + 10 + 5 + 1 = 32$$

where $C(5,0)$ corresponds to the case when there is no contextual information, $C(5,1)$ corresponds to the number of executions with one populated field, $C(5,2)$ with two populated fields, and so on. Table 10 summarizes the contextual information used in the executions that were run for the previous example. We executed the recommendation system for all fields in each instance of the test sets. In the previous example, the expected value for the disease field is *prostate cancer*, so we would compare the results of each of the 32 executions in Table 10 with that value.

Table 10. Contextual information used in each test run for a template instance with the following field–value pairs: *sex=male, organism part=prostate, cell line=PC-3, cell type=prostate cell, disease=prostate cancer, ethnicity=Caucasian*. The target field is *disease*. For each execution, the table shows the number of fields that constitute the context, as well as their names and values.

| Execution | Contextual information | |
|---|---|---|
| | No. fields | Field names and values |
| 1 | 0 | (no context) |
| 2 | 1 | *sex=male* |
| 3 | 1 | *organism part=prostate* |
| 4 | 1 | *cell line=PC-3* |
| 5 | 1 | *cell type=prostate cell* |
| 6 | 1 | *ethnicity=Caucasian* |
| 7 | 2 | *sex=male, organism part=prostate* |
| 8 | 2 | *sex=male, cell line=PC-3* |
| 9 | 2 | *sex=male, cell type=prostate cell* |
| ... | ... | ... |
| 32 | 5 | *sex=male, organism part=prostate, cell line=PC-3, cell type=prostate cell, ethnicity=Caucasian* |

## 4.7 Step 7: Analysis of results

The results of our evaluation for the eight experiments that we performed (see Table 8) are shown in Figure 9. The two plots on the top row show the results of single-template recommendations (experiments 1–4). The plots on the bottom row show the results of cross-template recommendations (experiments 5–8). All plots compare the performance of CEDAR's Value Recommender with the baseline, both for plain text values and for ontology terms, using different levels of contextual information. The x-axes of the plots in Figure 9 show the number of populated fields used as context to generate the recommendations. The y-axes show the mean reciprocal rank (MRR) obtained for each experiment. Table 11 shows the evaluation results by experiment and by number of populated fields in the context. Table 12 summarizes the evaluation results by type of recommendation algorithm (i.e., baseline, CEDAR's Value Recommender), recommendation scenario (i.e., single-template, cross-template), and type of metadata (i.e., text-based or ontology-based).

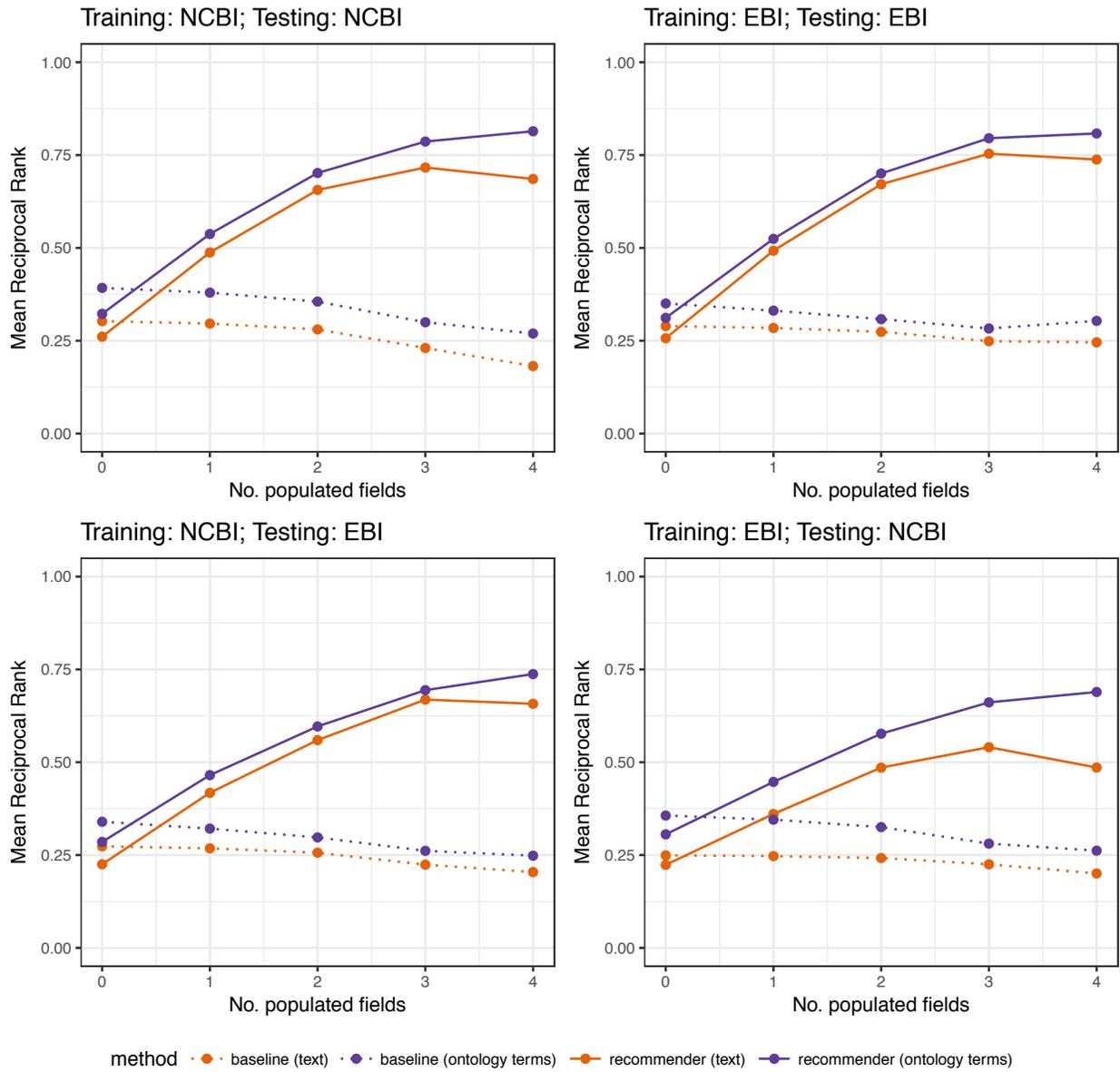

Figure 9. Mean reciprocal rank (MRR) for CEDAR's Value Recommender (*solid lines*) and for the baseline (*dotted lines*) using text-based metadata (*orange lines*) and ontology-based metadata (*purple lines*). The two plots on the top show the results of single-template recommendations (experiments 1-4). The two plots on the bottom row show the results of cross-template recommendations (experiments 5-8). The x-axis shows the number of populated fields used as context to generate the recommendations. The y-axis shows the MRR obtained for each experiment.

Table 11. Mean reciprocal rank (MRR) provided by the baseline and by CEDAR's Value Recommender for the 8 experiments in Table 8, specified by number of populated fields. Experiments 1–4 correspond to single-template recommendations, while experiments 5–8 correspond to cross-template recommendations. Experiments 1, 3, 5, 7 were performed using text-based metadata, while experiments 2, 4, 6, 8 were done using ontology-based metadata. NCBI: NCBI BioSample dataset. EBI: EBI BioSamples dataset.

| Experiment | Train/Test DB | Type of metadata | No. populated fields (baseline) | | | | | No. populated fields (Value Recommender) | | | | |
|---|---|---|---|---|---|---|---|---|---|---|---|---|
| | | | 0 | 1 | 2 | 3 | 4 | 0 | 1 | 2 | 3 | 4 |
| 1 | NCBI/NCBI | Text-based | 0.30 | 0.30 | 0.28 | 0.23 | 0.18 | 0.26 | 0.49 | 0.66 | 0.72 | 0.69 |
| 2 | NCBI/NCBI | Ontology-based | 0.39 | 0.38 | 0.36 | 0.30 | 0.27 | 0.32 | 0.54 | 0.70 | 0.79 | 0.81 |
| 3 | EBI/EBI | Text-based | 0.29 | 0.28 | 0.27 | 0.25 | 0.25 | 0.26 | 0.49 | 0.67 | 0.75 | 0.74 |
| 4 | EBI/EBI | Ontology-based | 0.35 | 0.33 | 0.31 | 0.28 | 0.30 | 0.31 | 0.52 | 0.70 | 0.80 | 0.81 |
| 5 | NCBI/EBI | Text-based | 0.27 | 0.27 | 0.26 | 0.22 | 0.20 | 0.23 | 0.42 | 0.56 | 0.67 | 0.66 |
| 6 | NCBI/EBI | Ontology-based | 0.34 | 0.32 | 0.30 | 0.26 | 0.25 | 0.29 | 0.47 | 0.60 | 0.69 | 0.74 |
| 7 | EBI/NCBI | Text-based | 0.25 | 0.25 | 0.24 | 0.23 | 0.20 | 0.22 | 0.36 | 0.49 | 0.54 | 0.49 |
| 8 | EBI/NCBI | Ontology-based | 0.36 | 0.35 | 0.33 | 0.28 | 0.26 | 0.31 | 0.45 | 0.58 | 0.66 | 0.69 |
| | | Mean (Text-based) | 0.28 | 0.28 | 0.26 | 0.23 | 0.21 | 0.24 | 0.44 | 0.60 | 0.67 | 0.65 |
| | | Mean (Ontology-based) | 0.36 | 0.35 | 0.33 | 0.28 | 0.27 | 0.31 | 0.50 | 0.65 | 0.74 | 0.76 |

Table 12. Mean reciprocal rank (MRR) provided by the baseline and by CEDAR's Value Recommender for the 8 experiments in Table 8, aggregated by recommendation scenario (i.e., single-template, cross-template) and by type of metadata (i.e., text-based or ontology-based).

| Method | Scenario | | Type of metadata | | Mean |
|---|---|---|---|---|---|
| | Single-template | Cross-template | Text-based | Ontology-based | |
| Baseline | 0.29 | 0.27 | 0.25 | 0.32 | **0.28** |
| Value Recommender | 0.60 | 0.51 | 0.52 | 0.59 | **0.56** |

By examining the results, we can clearly see that the MRR of the method improves as the amount of number of populated fields increases. This increase illustrates the strong influence of the context on the accuracy of the recommendations and demonstrates that our method is able to take advantage of contextual information to enhance the suggestions. The best results are obtained when using four populated fields as context, with an average MRR of 0.65 for textual metadata and 0.76 for ontology-based metadata. These results represent an improvement factor of 2.7 and 2.45 with respect to the results obtained without any contextual information (MRRs of 0.24 and 0.31, respectively).

The results obtained are consistently good for both single-template and for cross-template recommendations, demonstrating the ability of our approach to reuse metadata created from a template different than the template that the user is filling out. This is a key and novel feature of our framework that makes it possible to take advantage of existing metadata to generate recommendations for new templates without having to first populate the templates multiple times. Even though the results are good in both scenarios, the MRR is slightly better for single-database recommendations (0.60) than for cross-database recommendations (0.51). This is an expected behavior when generating cross-template recommendations, since the template instances in the training and testing datasets are generated using metadata from two different databases. Thus, the rules extracted from the training dataset may not cover all the cases in the test dataset. Our approach performs better for ontology-based metadata than for textual metadata for all the experiments, with an average MRR of 0.59 for ontology-based metadata and 0.53 for text-based metadata and. When using ontology-based metadata, it is possible to unify different terms that have the same meaning and therefore to overcome the limitations caused by term heterogeneity.

Finally, we analyzed the results for each field independently to determine whether the accuracy of the recommendations was affected by the specific target field used. Figure 10 shows the results for each field, for both our method and for the baseline, and for text-based metadata and ontology-based metadata. The x-axis shows the field name, and the y-axis the average MRR obtained for each field. The results are consistent with our previous findings and, interestingly, they also show that the highest improvement with respect to the baseline is obtained for those fields that have a large number of unique values in the training sets, such as *disease* and *cell line*. For those fields, the baseline statistic always returns the most frequent value and, since there are multiple possible values, the chances of failing are high. Since our approach uses the context to determine the most appropriate results, it is able to generate significantly better results than the baseline for fields with many possible values.

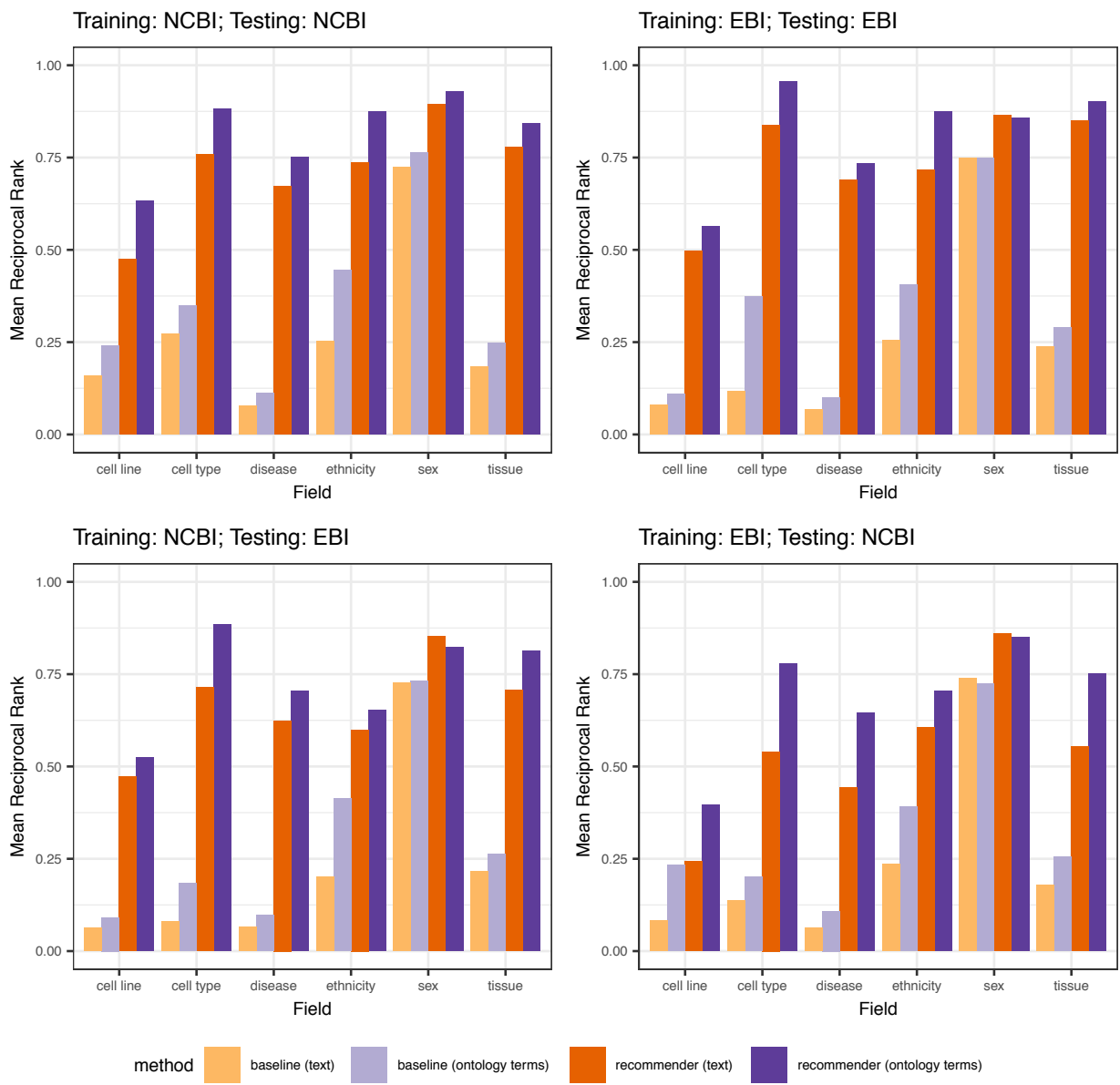

Figure 10. Mean reciprocal rank (MRR) provided by the baseline and by CEDAR's Value Recommender for the fields used in our evaluation (*cell line*, *cell type*, *disease*, *ethnicity*, *sex*, and *tissue*), using text-based metadata (*orange bars*) and ontology-based metadata (*purple bars*). The two plots on the top show the results of single-template recommendations (experiments 1–4). The two plots on the bottom row show the results of cross-template recommendations (experiments 5–8). The x-axes show the field name. The y-axes show the average MRR obtained for each field.

Our evaluation was focused on a subset of 6 fields commonly used to describe human samples. However, some users may need to use our system with a larger set of fields. We conducted an additional experiment to quantify the impact of including a larger number of fields on both the performance of the rule extraction process and on the accuracy of the suggestions provided. We used metadata from the NCBI BioSamples database to analyze the performance of the system for the 6 selected fields versus all the fields in the template (26 fields).[3] The results show that adding more fields to the evaluation considerable increased the number of rules generated and the time needed to generate the rules. The difference in the accuracy of the suggestions and in the time needed to generate them was minimal. The performance of our implementation for large datasets and with a large number of fields could be improved by replacing Apriori by a more efficient algorithm, such as FP-Growth (37).

## 5 Discussion

We have demonstrated that our framework successfully exploits associations between field–value pairs in existing templates to suggest field values in new templates. A key feature of the method is its use of populated fields during template completion to refine suggestions for unpopulated fields. The analysis of the results demonstrates that using this contextual information provides significant improvements in terms of the accuracy of the recommendations. Ontology-based mapping techniques also proved crucial for aligning fields and field values so that semantic associations could be detected and exploited by the recommendation method.

Our framework is an evolution of an earlier method that provided ontology-based, context-sensitive suggestions for metadata templates in the CEDAR Workbench (24). The previous method worked with values created using a single template only and could not learn from instances of templates that were structurally different. A key advantage of the new approach with respect to our previous work is that it does not require having any values previously entered for a given template. The new approach is effectively able to reuse metadata from other templates to generate suggestions.

A limitation of our approach is that appropriate domain ontologies must be available to provide standardized suggestions for values. As demonstrated in this paper, while a pure text-based approach can sometimes provide useful suggestions, performing associations at this level misses many lexically different but semantically identical values. The method also requires manual annotation of fields in templates with ontology terms to support cross-template alignment. A related shortcoming is that the framework supports only categorical field values; fields with continuous values (e.g., age) are currently ignored both by the learning and recommendation phases in the method.

---

[3] The results of this experiment are available in our Jupyter notebook (see "Additional experiment #2" at https://goo.gl/GtK956).

The evaluation was restricted to metadata about human samples from two public repositories. We plan to carry out deeper analyses of biological samples from other species to determine how our method generalizes to other sample types. The effect of mixing metadata about different organisms is expected to generate rules for a particular organism mixed with rules for other organisms. Some of those rules might be limited to organism-specific attributes, such as 'cultivar' and 'ecotype' for plants, but other rules can contain attributes that are valid for several organisms (e.g., disease). A potential problem is that the system may use a rule that was generated from metadata about an organism to generate suggestions for a different organism. The context-matching score allows the system to generate recommendations even when there is no perfect match between the antecedent of a rule and the context entered by the user and our evaluation has shown that, overall, this scoring approach provides good results. However, when mixing metadata from different organisms, the system might generate some suggestions that are biologically invalid. This negative effect can be controlled by either setting a higher threshold for the final recommendation score (e.g., 0.8); or by cancelling the influence of the context matching score factor in the final recommendation score, so that the system will only take into account the rules whose field-value pairs in the antecedent match perfectly the field-value pairs entered by the user.

Future efforts will also concentrate on evaluating the framework with a larger number of repositories to see how more complex relationships can be discovered and integrated to enhance the suggestions generated by the system. We plan to conduct a user-based evaluation in collaboration with the Adaptive Immune Receptor Repertoire (AIRR) Community (https://www.antibodysociety.org/the-airr-community) to assess if the introduction of our method is actually bringing an advantage to biomedical scientists in terms of facilitating and speeding up metadata entry. AIRR researchers identified the lack of standards to describe their datasets as a bottleneck to their progress and produced a metadata standard, known as MiAIRR (38), for capturing the principal characteristics of experiment types, collectively referred to as repertoire sequencing. In collaboration with members of the AIRR community, we operationalized an end-to-end submission pipeline (39) based on a CEDAR template known as MiAIRR, which scientists can use to enter metadata associated to AIRR studies and to upload metadata and associated sequencing data to the three target NCBI repositories: BioProject, BioSample, and SRA. Our plan is to enable metadata suggestions for the BioSample section of the MiAIRR template using metadata from the NCBI BioSample and EBI BioSamples databases and perform a user-based evaluation to assess the usefulness of our recommendation approach.

In addition to facilitating the process of entering new values for metadata templates, our work also has implications for retrospective augmentation of existing metadata. For example, the framework could be used to identify potential mistakes in previously entered values by applying it not only to empty fields, but also to populated fields. Strong differences between the existing value for a field and the values recommended by the system could be highlighted to human curators for review. The framework could also be used to generate suggestions for multiple fields at a time and use suggestions with a recommendation score higher than a predefined threshold to automatically populate those fields.

# 6 Conclusion

Despite the importance of metadata to facilitate data discovery, interpretation, and reuse, metadata for online datasets are generally of poor quality. A key problem is that the typical metadata acquisition process is tedious and time consuming, with little or no support provided to users. In this paper, we described a method that uses association rule mining coupled with ontology-based semantic mappings to provide suggestions to users creating metadata. We have implemented this method as a Web service and integrated it into the CEDAR Workbench, an end-to-end platform for metadata authoring and submission. The resulting service is known as CEDAR's Value Recommender.

Our approach takes advantage of associations among values in existing metadata to generate context-sensitive recommendations when creating new metadata. A key focus is on interoperation with ontologies. This interoperation has the dual aim of (1) aligning text-based values with ontology terms to support high level analysis, and (2) providing ontology-based value suggestions to users to increase standardization of the entered values. A novelty of the method is that it can discover associations in existing instances of structurally different templates and use those associations to provide context-sensitive recommendations for new templates. While the driving impetus of CEDAR's Value Recommender is to help biomedical investigators to quickly annotate their experimental data with metadata, the approach can be applied to any domain for which a suitable set of domain ontologies is available.

We evaluated our approach using metadata from the NCBI BioSample and EBI BioSamples databases. The evaluation focused on determining the effectiveness of the method for generating metadata suggestions for both repositories. The results suggest that our method has the potential to help investigators easily and quickly create comprehensive and standardized metadata for their experimental datasets, thus increasing adherence of the datasets to the FAIR principles in support of open science.

## Acknowledgements

CEDAR is supported by the National Institutes of Health through the NIH Big Data to Knowledge program under grant 1U54AI117925. NCBO is supported by the NIH Common Fund under grant U54HG004028. The authors thank Rafael Gonçalves for his valuable suggestions to improve the manuscript. All software described in this paper is open source and available on GitHub (https://github.com/metadatacenter).

## Availability of software and data

We have created a Jupyter notebook describing in detail the steps to reproduce our evaluation using Python and R scripts. The notebook contains links to the datasets and software used, as well as to the rules and results generated. The notebook is available at https://github.com/metadatacenter/cedar-experiments-valuerecommender2019/blob/master/ValueRecommenderEvaluation.ipynb. CEDAR's Value Recommender source code is available at https://github.com/metadatacenter/cedar-valuerecommender-server. Documentation on how to use the CEDAR Workbench to enable recommendations for CEDAR templates is available at https://github.com/metadatacenter/cedar-docs/wiki/CEDAR-Value-Recommender.

receptor repertoire sequencing studies to the NCBI. Front Immunol. 2018;9:1877. 10.3389/fimmu.2018.0187748